\documentstyle[11pt,paspconf]{article}

\begin{document}
\input{psfig}

\title{A critical look at the role of AGB stars in stellar population 
synthesis}
\author{A. Lan\c{c}on}
\affil{Observatoire astronomique, Universit\'e L. Pasteur \& CNRS, UMR\,7550,
11 rue de l'Universit\'e, 67\,000 Strasbourg, France}

\begin{abstract}
Asymptotic giant branch stars are essential contributors to the near and
mid-IR emission of intermediate age (10$^8$-10$^9$\,yr old) stellar 
populations. Detecting this light will set constraints on the star
formation history in galaxies and, conversely, the search for AGB
signatures in well studied populations will help us reduce some of
the still large uncertainties in AGB models. This paper reviews
how AGB stars are currently included in population synthesis models and
which spectral features can be used to identify their emission in
galaxy light; targets for observational tests are suggested, and some
observational and theoretical difficulties are discussed. 
\end{abstract}


\section{Introduction}
Asymptotic Giant Branch stars (AGB stars)
are rare members of stellar populations. However, they are 
among the most luminous cool stars and can therefore be detected sometimes
even individually in external galaxies; they are also the coolest of the 
luminous stars and hence produce spectral signatures that we may search 
for in integrated near/mid-IR spectra. Finally, the carbon we are made of 
and the pre-solar meteorites found on Earth remind us of the role played
by AGB stars in the evolution of galaxy matter.

Population synthesis techniques predict the evolution of stellar distributions
in the HR diagram and the corresponding integrated emission properties.
Nowadays they also keep track of the chemical evolution. 

In this review, we will focus on the spectrophotometric 
aspect of population synthesis, at solar and LMC metallicities. We will
discuss how we can predict the contribution of AGB stars to the integrated
spectrum and how we can recognise these contributions in observed spectra. Much of this work has become feasible only recently thanks to advances in 
AGB theory and to more complete samples of AGB observations.
We will then suggest extragalactic targets and discuss what the 
detection -- or not -- of AGB signatures can tell us about the 
galaxies or alternatively about AGB evolution. The relevant observations are 
becoming feasible now and are mostly work of the future.

\section{The contribution of AGB stars to the bolometric luminosity}

Predicting the contribution $f_{\rm{bol}}$ of AGB stars to the bolometric
luminosity as a function of time for an evolving stellar population (SP)
``only" requires a set of stellar evolution tracks. Such tracks 
used to be constructed semi-empirically (e.g. Charlot \& Bruzual 1991)
and are now provided by so-called synthetic AGB evolution models
(Groenewegen \& de Jong, 1993, Marigo, this conference). The 
effects of thermal pulses are taken into account in these calculations,
but the product used as input in population synthesis calculations is
a set of smooth tracks that needs to be well understood as a 
representation of either the interpulse properties or some mean properties,
e.g. the energy weighted averages over helium shell flash cycles. The
precise definition determines further steps in the synthesis code. In the 
first case for instance, corrections should be applied to account for
the post-flash luminosity dip.\footnote{Synthetic evolution was first
introduced as a means of simplifying population synthesis calculations;
we note however that synthetic models become more and more complex,
and that a return to the direct use of the output of full evolutionary 
calculations may become attractive again.}

The critical parameter in setting $f_{\rm{bol}}$ is the mass loss $\dot{M}$,
which determines AGB lifetimes and the range of initial masses for which the 
AGB phase occurs. Other parameters are implicit in $\dot{M}$: e.g., metallicity 
influences grain formation and the efficiency of radiation pressure, the 
mixing parameters and gas opacities influence the extension of the stellar 
envelopes and the Mira-type pulsation properties, that contribute to
starting the winds. Available tracks usually differ in more than one of 
these parameters, which makes it difficult to interpret differences in 
lifetimes and energy distributions; at this stage, the scatter among
the lifetimes found in the litterature for the thermally pulsating 
AGB (TPAGB) is about a factor of 2 and must be treated as an 
uncertainty.

The predictions reported in this paper are based on the population
synthesis code of Fioc \& Rocca-Volmerange (hereafter FRV;
Z=0.02: 1997, other Z: 1998).
A Salpeter IMF with a lower mass limit at 0.1\,M$_{\odot}$ is adopted 
in all calculations. The TP-AGB evolution is included in the code
following the prescriptions of Groenewegen et al. (1993, 1995); the
resulting tracks follow those of Vassiliadis \& Wood (1993) closely
in the HR diagram. The FRV lifetimes are always close to the shortest ones
found in the litterature, therefore we do not expect to overestimate
the effect of the AGB. A more complete study will be reported elsewhere.

After an instantaneous burst of star formation (IB), we find that the
contribution $f_{\rm{bol}}$ due solely to the TPAGB stars reaches maximal values
($\ga 10\,\%$) between {\bf 0.3 and $\sim 1.5$\,Gyr}, corresponding
to AGB progenitor masses between 1.7 and 3.5\,M$_{\odot}$. At 1.5\,Gyr
the complete AGB (EAGB+TPAGB) contributes about a quarter of the light; values
of up to 40\,\% have been reported (Charlot \& Bruzual 1993). At
later stages, the light of the TPAGB stars is progressively lost in the 
increasing contribution of EAGB and first giant branch stars.

AGB lifetimes can in principle be tested with integrated V-K colours since 
the AGB contribution is essentially emitted at near-IR wavelengths. The
TPAGB by itself is responsible for more than 40\,\% (and up to 60\,\%) 
of the K band light {\bf 0.1 to 1\,Gyr} after an IB. 
At these ages, the red supergiants
responsible for very red colours earlier on have
died. We refer to Charlot, 1996, for an interesting model to model
comparison of colour and AGB contribution predictions.

Girardi \& Bertelli (1998) show how changes 
in $\dot{M}$ affect V-K; in particular, the increased $\dot{M}$ and hence
shorter lifetimes resulting from envelope burning in massive AGB stars
may reduce V-K by up to 1~magnitude at the earliest AGB-dominated times.
Some ambiguities and practical difficulties faced when performing tests 
based on colours are discussed in Sect.\,\ref{res.sect} and \ref{obs.sect}

\section{The effect of AGB stars on the near and mid-IR spectrum}

\subsection{Constructing a suitable stellar library}
 
In order to predict the distribution of AGB light along the IR spectrum and
to determine how AGB stars can be recognised among other cool
contributors (red supergiants and first giant branch stars), 
stellar spectroscopic
libraries are needed. Commonly used near-IR libraries include the
empirical ones of Kleinmann \& Hall (1986), Terndrup et al. (1991),
Lan\c{c}on \& Rocca-Volmerange (1992), the theoretical one of 
Bessell et al. (1991) and the semi-empirical compilation of Lejeune et al.
(1998). None of these consider the pulsating long period variables
(LPVs) explicitely, despite the fact that pulsation is known to
modify the spectral type\,--\,temperature\,--\,colour relations and to increase
molecular absorption features (Johnson \& Mendez, 1970, Bessell et al. 1989),
and that basically all the coolest AGB stars are LPVs. As a consequence,
near-IR stellar energy distributions (SEDs) of galaxies have been interpreted
in terms of a supergiant and giant dichotomy (mainly on the basis of CO 
absorption at 2.3\,$\mu$m and the Ca triplet at 850\,nm), 
but no segregation among cool giant populations has been possible.

Lan\c{c}on \& Wood (1997, hereafter LW97) 
have obtained 0.5-2.5\,$\mu$m spectra for
$\sim$100 cool stars, including instantaneous spectra of a sample of 
O-rich and C-rich LPVs observed at various phases.
The observations confirm that only LPVs are able to produce the 
deepest H$_2$O bands around 1.4 and 1.9\,$\mu$m
(Mouhcine \& Lan\c{c}on 1998). Other characteristic
near-IR molecular bands are those of VO (1.05\,$\mu$m) and TiO (1.25\,$\mu$m,
in the latest spectral types), or of CN (1.1, 1.4\,$\mu$m)
and C$_2$ (1.77\,$\mu$m)
in C-stars.

Temperatures were assigned to individual Mira-type spectra of LW97 in two ways.
On one hand, R.\,Alvarez compared the energy distribution (near-IR H$_2$O bands
excluded) with the static giant
models of Plez (model version of Alvarez \& Plez, 1998).
On the other, colour-temperature relations based on angular diameter
measurements were used (Feast et al. 1996). Below 3000\,K, the first 
temperatures are systematically higher than the second, the difference 
increasing up to $\sim$400\,K around 2000\,K. 
The LPV atmosphere models of Bessell et 
al. 1996 also indicate low temperatures (in the rather rare cases where the
fit is satisfactory). The two latter methods are not independent: angular
diameter interpretations require model atmospheres. In view of the
difficult reconciliation between large observed radii and fundamental mode
pulsation (Wood, this conference), there may be systematic errors
related to the use of current AGB atmosphere models and the second
temperature estimates may be too cool.
Therefore, the results presented here are based on an intermediate
scale.

\begin{figure}
\centerline{\psfig{figure=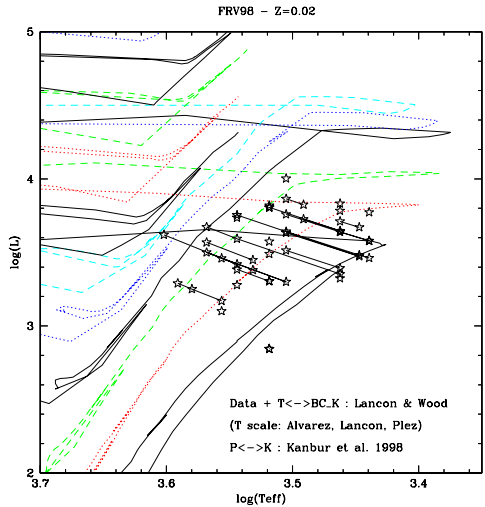,height=7cm,clip=}}
\caption[]{Plotted over the evolution tracks of FRV (1998) are the
locations of observed spectra from LW97. T$_{\rm eff}$ is
estimated according to the warmer scale described in the text; luminosities
are derived from a P-L$_K$ relation 
and the K band bolometric corrections are measured on the instantaneous
spectra themselves (K band variability is neglected). 
Lines connect observations of single stars at 
various phases.}
\label{tracks.fig}
\end{figure}

The variations of individual stars are illustrated in Fig.\,\ref{tracks.fig}.
Averaging the spectra star by star is not a practical way of obtaining a 
library for population synthesis: it requires complete phase
coverage for each star as well as an estimate of
its fundamental parameters (initial mass,
metallicity, evolutionary status). 
The latter are so uncertain that the resulting mean spectra are 
difficult to order. First attempts with the LW97 library
have shown that the expected global spectral evolution from higher to lower 
temperature is lost in object to object variations. It is more convenient 
to group the spectra in instantaneous temperature bins. Although the 
dispersion in the spectral properties inside each bin remains to be explained
(Mouhcine \& Lan\c{c}on, in preparation), this method provides a 
smoothly evolving sequence.

For population synthesis purposes, the LW97 spectra were combined with 
library of Lejeune et al. (1998), as implemented by FRV (1998).

\subsection{Results}
\label{res.sect}
As mentioned above, the AGB spectra we are able to distinguish from those
of older giants are those of {\em pulsating} stars. 

\begin{figure}
\centerline{\psfig{figure=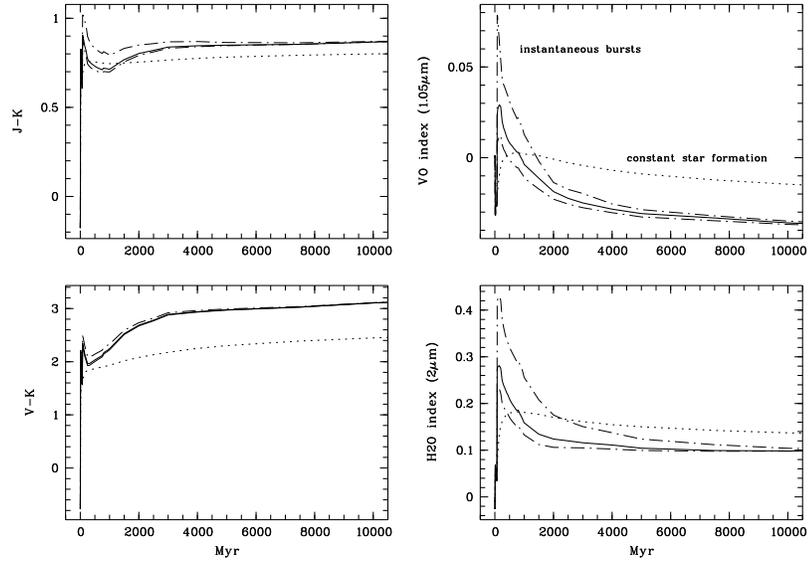,angle=270,height=8cm,clip=}}
\caption{Evolution of selected properties with time. The 
dot-dashed lines illustrate the burst evolution for the two extreme
spectral temperature scales discussed in the text. Molecular indices
are in magnitudes.}
\label{evol.fig}
\end{figure}

\begin{figure}
\centerline{\psfig{figure=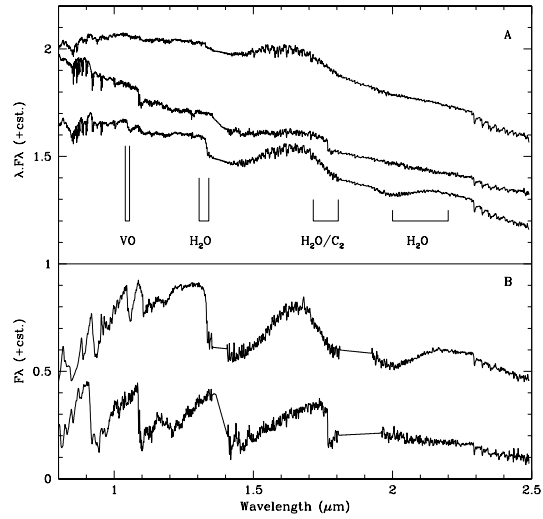,height=7.5cm,clip=}}
\caption{A: Selected synthetic integrated spectra. From top to bottom: 
a static giant dominated 4\,Gyr old population, and two 200\,Myr old postbursts,
when most TPAGB stars are respectively C-rich and O-rich. Filter
centers for useful molecular indices are indicated. 
B: Typical cool O-rich and C-rich LPV spectra.}
\label{bandids.fig}
\end{figure}

Fig.\,\ref{evol.fig} 
shows the predicted behaviour of colours and molecular indices 
at solar metallicity, with the 
assumption that the TPAGB and LPV stages coincide and that LPVs remain
visible and O-rich along most of the TPAGB. The blue post-starburst colours 
are due to the inevitable contamination by luminous intermediate
mass blue main sequence stars, coeval with the AGB population.

Several parameters are degenerate in studies based on colours
alone: extinction, metallicity and age all redden the integrated spectra,
continuous star formation keeps them blue over long times.
As an example, the ratio of an IB spectrum at 500\,Myr to the spectrum of a 
constantly star forming population at 2\,Gyr is essentially flat
between 0.5 and 2.5\,$\mu$m. Spectral features must be observed to 
distinguish the two. As we know from previous models and from
observations of star clusters, the Balmer and Paschen series of 
numerous A type stars characterise the younger of the two populations 
below 1\,$\mu$m.  When LPV spectra are included in the models, 
H$_2$O bands appear in the near-IR part of the ratio
as broad absorption features at the same relative level as the H$\alpha$ line
(see also Fig.\,\ref{bandids.fig}\,A).

The right hand side of Fig.\,\ref{evol.fig}
demonstrates that the molecular bands of 
H$_2$O and of VO (1.05\,$\mu$m) unambiguously characterise 
``postburst populations" 0.1 to 1\,Gyr after an IB, with a peak
at 150\,Myr. The maximal numerical values of the molecular 
indices however depend strongly on the adopted T$_{\rm eff}$ scale. 
In the case of a constant star formation rate,
only the warmer T$_{\rm eff}$ scale (which leads to
the use of spectra with deeper features at a given point of an
isochrone) leads to a significant increase
of the molecular bands, when compared to the predictions of FRV (1998) 
without the specific LPV library.

These results remain qualitatively valid in LMC metallicity models.

\subsection{The effects of various model assumptions}

\hspace*{\parindent}
3.3.1 -- Does the TPAGB coincide with the instability strip for LPV pulsation?
Period-luminosity (P-L) diagrams of observed Miras
suggest that the instability strip is narrower. If we transport the blue
edge of the strip into the population synthesis code using the prescription 
of Vassiliadis \& Wood (1993), we find that the effect on the colours is
negligible (LPV spectra are replaced by static giant spectra with
very similar energy distributions) but that the depth of the H$_2$O
features is reduced during the 2\,Gyr following an IB (by $\sim0.03$\,mag
for the index in Fig.\,\ref{evol.fig}). 
VO is not affected: it is produced by the cooler Miras.

In fact, recent data supports a more complex picture (Alvarez et al. 1997,
Barth\`es 1998, Wood, this conference, Kiss, this conference), 
that also shows through in 
theory (Fox \& Wood 1982, Xiong et al. 1998, Tuchman, this conference). 
Pulsation may be intermittent, depending on phase in the thermal pulse 
cycle; but globally, AGB stars start varying as small amplitude high
overtone pulsators and successively (with overlaps in time)
become unstable to lower mode, higher amplitude and longer period
pulsation. The Miras of the empirical 
P-L relations are only the largest optical amplitude
pulsators. Since we have not excluded low amplitude LPVs from our 
spectral temperature bins, these bins represent a mixture of 
pulsation states that may reasonably approximate reality
in models based on coinciding TPAGB and LPV stages.

At the red edge of the LPV instability strip, stars become invisible
even in the near-IR due to ejected material, and instead emit longwards of
3\,$\mu$m. Bressan et al. (1998) studied the consequences on integrated
broad band colours: during the AGB-dominated post-starburst stages,
the optical/near-IR spectrum becomes bluer due to the missing late
AGB contributions, while the K-$\lambda$ colours with $\lambda >5\,\mu$m
increase dramatically due to OH/IR type infrared sources. 
At LMC metallicities, where the red edge is best
defined, we find that V-K changes by a few percent and 
the molecular features are hardly affected at all. The effect could be stronger
at higher metallicities.

3.3.2 -- Some AGB stars will end their lives as C stars, with completely 
modified spectra. An exploratory study based on the relative
C-rich TPAGB lifetimes of Groenewegen \& de Jong (1993),
at the metallicity of the LMC, shows that this totally removes any oxide 
features appart from CO for most of the duration of the AGB-dominated
post-starburst. Fortunately, the C$_2$ bandhead at 1.77\,$\mu$m
then replaces the H$_2$O features, and a narrow band photometric index can 
be defined that measures the time dependent AGB contribution, 
whether the AGB stars be O-rich or C-rich (Fig.\,\ref{bandids.fig};
Lan\c{c}on et al., 1998).

\section{Observational perspectives: where, how and why?}
\label{obs.sect}

\subsection{Observational strategies}
The presence of pulsating AGB stars modifies the near-IR spectral signatures
and the mid-IR colours during several Gyr after an episode of enhanced
star formation and more permanently if star formation continues.
However, at the time of this conference the uncertainties on
fundamental stellar inputs such as the Mira temperature scale (necessary
to connect suitably averaged stellar spectra to evolutionary tracks), 
the mass-dependent TP-AGB and the LPV lifetimes, 
the O-star to C-star transition, limit the number of {\em robust }
predictions directly applicable to extragalactic studies to a few: we
expect unambiguous AGB signatures 0.1 to 1\,Gyr after a starburst, 
and we know  we may search them using combined near and mid-IR
colours (Bressan et al. 1998) or near-IR molecular features
(Lan\c{c}on et al. 1998). Other diagnostic features will certainly 
appear as soon as the astronomical community will have
digested the ISO data. It is important that the most robust predictions
be tested observationnally before more difficult data interpretations 
(mixed populations with a complex star formation history) be undertaken.

Significant populations with appropriate ages exist in
star clusters of Local Group galaxies, in the bodies of some of these
galaxies, in E+A galaxies and in post-starburst galaxies.
\smallskip

Star clusters, in particular the LMC age sequence, may look like ideal targets
but each of them contains only a handful of luminous AGB stars. 
The integrated colours are significantly affected by the stochastic 
fluctuations in the AGB numbers (Ferraro et al. 1995, Santos \& Frogel 1997,
Marigo et al. 1996). 
Assuming Poisson statistics for these small numbers, we can
write the relative, time and wavelength dependent luminosity fluctuations
due to TPAGB stars analytically as:
$$ \frac{\sigma_L }{L} =
\frac{1}{\sqrt{N}} . \frac{\sqrt{\alpha} \,(L_a
     - L_r )}{\alpha \,L_a + (1-\alpha)\,L_r } $$
Here, $N$ is the total number of stars in the population,
$\alpha$ is the (small) proportion of TPAGB stars, $L_a (\lambda,t)$ and
$L_r (\lambda,t)$ are respectively the average emissions per TPAGB or
non-TPAGB star and are both obtained from population synthesis 
calculations. With the above mentioned IMF, more than $10^6$
stars are required in order to keep these fluctuations below 10\,\%
at K a few 100\,Myr after an IB. They will then be 10 times smaller
at 0.7\,$\mu$m or in terms of bolometric luminosities, and become
negligible at K within a Gyr. 

Nevertheless, now that the AGB numbers in individual Local Group star clusters
are known and their effect on the colours has been demonstrated
it remains interesting to test whether their near-IR spectroscopic 
features show through. This requires either several slit positions per
cluster, or wide field or imaging spectroscopy.
\smallskip

Larger AGB numbers are present in the bodies of selected nearby dwarf
or E+A galaxies, but they are superimposed on older populations and
their spectral signatures will be severely diluted. If we adopt
the maximal value reached by a molecular absorption index in 
a constant star formation scenario as the threshold for the unambiguous
detection of AGB contributions (Fig.\,\ref{evol.fig}), 
we find that postburst populations
can be identified as long as they represent more than $\sim$10\,\% of the 
total stellar mass in the observed area (Lan\c{c}on et al. 1998). This
result is relatively independent of the SF history of the 
supposedly old underlying population.

Counts should be used preferentially in the studies of Local Group
galaxies (e.g. Gallart, 1998), 
especially since integrated near-IR spectra are still
difficult to obtain for extended objects that don't have the 
high surface brightnesses of currently star forming regions. One of the
main observational difficulties is crowding (Martinez-Delgado \& Aparicio,
1997). It may also be possible to study AGB contributions with extended
``pixel fluctuation methods" (as used in microlens searches, e.g.
AGAPE, Ansari et al. 1997)\footnote{An idea first mentioned to
me by A.L.\,Melchior}, when pixels integrate the light of less than
$10^6$ stars.
\smallskip

It follows from the above arguments that the
ideal targets for first tests should be {\em both massive and localised,
0.1-1\,Gyr old postburst populations}. 
A few such objects have recently been identified:
new clusters with estimated stellar masses of up to 
$10^7$-$10^8$\,M$_{\odot}$ form in interacting galaxies and can
apparently survive as bound objects for more than the 0.1\,Gyr
required for AGB stars to appear. Clusters with appropriate 
estimated ages are listed for NGC\,7252, NGC\,1275 and NGC\,3921 
(Miller et al. 1997 and references therein)\footnote{
A proposal has been submitted to
perform a near-IR search for the predicted AGB signatures}. 
Neither dilution nor statistical fluctuations are expected to represent
serious hurdles there. Many of the more extended (but still bound)
post starburst areas predicted by the dynamical 
arguments of Kroupa (1998) would also satisfy the detection criteria.

Before closing this section, we wish to mention observational difficulties.
The IR signatures of interest are located near the strong telluric
absorption bands of H$_2$O and CO$_2$, calling for particularly frequent
observations of a calibration star during ground based observations. 
High signal to noise ratios per (low resolution) spectral element are
required to detect molecular absorption bands. On the other hand,
dust emission associated with ongoing star formation may hinder 
mid-IR searches for OH/IR emission in places. Spatial resolution must
always be sought to limit dilution. 

\subsection{Perspectives}

{\bf What can we learn about galaxies?} \\
Once the first tests on populations with
known ages will have confirmed the predictions regarding the IR signatures
of AGB-rich populations, these observations will be useful in the
study of the star formation history of galaxies, the propagation
of star formation, the survival of starburst clusters and the
potential precursors of  globular clusters or dwarf galaxies. Recent
star formation is often associated with extinction, in which case 
optical (e.g. Balmer line) studies present incomplete pictures.
Post-starburst knots may be found that coexist with current starbursts in 
dusty, infrared lumionous objects like NGC\,253, IC\,342.

In more evolved objects, late stellar evolution paradoxally helps us
understand the UV emission: UV-excesses, previously interpreted as
signatures of very recent star formation, are now explained by relatively
old AGB-manqu\'es or Hot Horizontal Branch objects, closely related 
to the stars discussed in this conference (Bressan et al. 1994,
Brown et al. 1995, FRV 1998).

Finally, as suggested by Bressan et al. (1998), AGB stars may be used to break
the well-known age-metallicity degeneracy of galaxy colours. 
In brief, the red giant population
of galaxies increases with time and with metallicity, leading to similarly
red integrated colours in both cases; but while the mid-IR emission of
OH-IR stars in galaxies is predicted to increase with metal abundance, it
decreases with age. A similar argument holds for the near-IR
molecular TP-AGB signatures, although the predicted effect is small.

\smallskip

\noindent{\bf What can we learn about the AGB?}\\
The tests  discussed in Sect.\,4.1 will provide constraints on
AGB lifetimes (i.e. on stellar evolution models), on temperature
scales (i.e. on pulsating atmosphere models) and on the O- to C-star
transition (i.e. on dredge-up). All these parameters badly need to 
be better determined. 

The unique access to distant populations
provided by integrated spectroscopic observations will then
allow us to explore AGB evolution in more extreme environments.
For instance, it has been claimed that stars with ages of $\sim0.1$\,Gyr
dominate in the emission of the companion of quasar PG\,1700$+$518
at redshift 0.29 (Canalizo \& Stockton, 1997). 
Are AGB stars numerous there? Are they carbon rich or oxygen rich?
While more sensitive infrared instruments are
being constructed for high spatial resolution telescopes, we have
time to exploit Local Universe data and to improve upon existing AGB model
predictions.

\acknowledgments
It is a pleasure to acknowledge the active collaboration of
R.\,Alvarez, M.\,Fioc, M.A.T.\,Groenewegen, M.\,Mouhcine, B.\,Rocca-Volmerange,
M.\,Scholz, D.R.\,Silva and P.R.\,Wood in making this review possible.

\end{document}